\documentclass[a4paper, pra, twocolumn,superscriptaddress,showpacs,english]{revtex4-1}
\usepackage[T1]{fontenc}
\usepackage[latin9]{inputenc}
\usepackage{float}
\usepackage{amssymb}
\usepackage{graphicx}
\usepackage{babel}
\usepackage{color}
\usepackage[colorlinks,citecolor=blue,linkcolor=blue]{hyperref}
\begin{document}

\title{Spin-Orbit Coupling Induced Spin Squeezing in Three-Component Bose Gases}

\author{X. Y. Huang}
\affiliation{State Key Laboratory of Mesoscopic Physics, School of Physics, Peking
University, Collaborative Innovation Center of Quantum Matter, Beijing 100871, China}
\author{F. X. Sun}
\affiliation{State Key Laboratory of Mesoscopic Physics, School of Physics, Peking
University, Collaborative Innovation Center of Quantum Matter, Beijing 100871, China}
\author{W. Zhang}
\email{wzhangl@ruc.edu.cn}
\affiliation{Department of Physics, Renmin University of China, Beijing 100872, China}
\affiliation{Beijing Key Laboratory of Opto-electronic Functional Materials and Micro-nano Devices,
Renmin University of China, Beijing 100872, China}
\author{Q. Y. He}
\email{qiongyihe@pku.edu.cn}
\affiliation{State Key Laboratory of Mesoscopic Physics, School of Physics, Peking
University, Collaborative Innovation Center of Quantum Matter, Beijing 100871, China}
\affiliation{Collaborative Innovation Center of Extreme Optics, Shanxi University, Taiyuan, Shanxi 030006, China}
\author{C. P. Sun}
\email{cpsun@csrc.ac.cn}
\affiliation{Beijing Computational Science Research Center, Beijing 100084, China}

\begin{abstract}
\vspace{0.1em}
\noindent We observe spin squeezing in three-component Bose gases where 
all three hyperfine states are coupled by synthetic spin-orbit coupling. This 
phenomenon is a direct consequence of spin-orbit coupling, as can
be seen clearly from an effective spin Hamiltonian. By solving  this effective 
model analytically with the aid of a Holstein-Primakoff transformation for spin-1
system in the low excitation limit, we conclude that the spin-nematic squeezing,
a novel category of spin squeezing existing exclusively in large spin systems, 
is enhanced with increasing spin-orbit intensity and effective Zeeman field, which 
correspond to Rabi frequency $\Omega_R$ and two-photon detuning $\delta$
within the Raman scheme for synthetic spin-orbit coupling, respectively. These 
trends of dependence are in clear contrast to spin-orbit coupling induced spin 
squeezing in spin-1/2 systems. We also analyze the effects of harmonic trap 
and interaction with realistic experimental parameters numerically, and find that 
a strong harmonic trap favors spin-nematic squeezing. We further show spin-nematic
squeezing can be interpreted as two-mode entanglement or two-spin squeezing 
at low excitation. Our findings can be observed in $^{87}$Rb gases with existing 
techniques of synthetic spin-orbit coupling and spin-selectively imaging.
\end{abstract}
\maketitle

\section{Introduction}

Spin squeezing is an important resource which has many potential applications not only in quantum metrology and atom interferometers~\cite{Wineland-94, Kitagaw-93, Gross-08, Riedel-10, Gross-10}, but also in many aspects of quantum information due to its close relation with quantum entanglement~\cite{Hofmann-03, Knapp-09, Drummond-09, Reid-09, Cavalcanti-09}. In conventional experiments, squeezing is usually achieved via the nonlinearity induced by the inter-particle interaction~\cite{Gross-08, Riedel-10, Gross-10}. As an example, spin squeezing has been obtained experimentally in a Bose-Einstein condensate (BEC) of a three-component Bose gas~\cite{chapman-exp}. However, the intensity of spin squeezing in these experiments crucially depends on the interaction between atoms. In cold atom experiments, the background interaction is usually very weak such that the observation of squeezing is relatively hard. Although there are some techniques to enhance the interaction, e.g., by tuning the state-dependent microwave potentials~\cite{Riedel-10}, or through a magnetic Feshbach resonance in alkali atoms~\cite{Gross-10, Chin-10}, the side effects of decoherence, severe atom loss and dynamical instability induced by strong interaction still hinder the achievement of strong spin squeezing.

The experimental realization of synthetic spin-orbit coupling (SOC) in ultracold atomic gases~\cite{Spielman, Jing-Zhang, Zwierlein} has attracted much attention, partly due to its close relation to exotic many-particle states and novel excitations~\cite{Hui-Zhai2, Yirev, Jing-Zhang4}. Recently, theoretical studies have proposed to realize spin squeezing in two-component BEC by synthetic spin-orbit coupling (SOC)~\cite{chen-13, huang-15}. It has been shown that the presence of SOC will induce an effective spin-spin interaction which can lead to spin squeezing. However, there are two disadvantages of these proposals. First, the synthetic SOC requires a Raman transition between two hyperfine states. The Rabi frequency of this Raman transition is detrimental to spin squeezing, i.e., a stronger SOC leads to a weaker squeezing. Besides, the two-photon detuning of this Raman transition is also unfavorable such that best squeezing will be achieved when the detuning is zero. Nonetheless, in realistic experiments one would encounter severe heating effect when the detuning is tuned on resonance.

In this paper, we study spin squeezing in a three-component Bose gases where all three hyperfine states are coupled by spin-orbit coupling induced by Raman transitions. As a result, this system has pseudo-spin-1, and the spin operators herein must be described by SU(3) spin matrices, i.e., the Gell-Mann matrices. These Gell-Mann matrices span an eight-dimensional spin hyper-space, with three of them are usually refereed as spin vectors, and the other five as nematic tensors~\cite{Yukawa-13}. The squeezed spin operators hence can be categorized into three types, including the spin-spin squeezing, nematic-nematic squeezing, and spin-nematic squeezing. Here, we focus on the spin-nematic squeezing, as it is a novel type of squeezing which exists exclusively in systems with large spins. We find that the presence of SOC can induce spin-nematic squeezing, which can be further enhanced by increasing the SOC intensity or reducing the quadratic Zeeman splitting. These trends of dependence can be understood from an effective Hamiltonian, in which the Rabi frequency and quadratic Zeeman splitting correspond to effective Zeeman fields in the spin and nematic sectors, respectively, hence causing opposite effects on various types of spin squeezing. More importantly, we find that the squeezing is favored by two-photon detuning of the Raman transition within a fairly large parameter regime, which is beneficial for experimental realizations to avoid severe heating effect. When the system exhibits spin-nematic squeezing in the low excitation limit, we also find two-mode entanglement ~\cite{duan-02}  and two-spin squeezing ~\cite{lyou-02} in the system. We further study the effects of an external trapping potential and inter-atomic interaction which are present in realistic experimental situations by numerically analysis, and conclude that the spin-nematic squeezing is favored by stronger trapping potentials. Finally, we discuss possible detection scheme via a spin-selective imaging technique and a radio-frequency (RF) rotation of the spin axes~\cite{chapman-exp}.

The remainder of this paper is organized as follows. In Sec.~\ref{sec:formalism}, we introduce the system under investigation and discuss the single-particle spectra. We then derive an effective spin Hamiltonian from which it can be seen clearly that SOC induces an effective spin-spin interaction. We then analyze the spin-nematic squeezing and its dependence of various factors in Sec.~\ref{sec:results}. Finally, we discuss possible experimental detection scheme and summarize in Sec.~\ref{sec:conclusion}. 

\section{single-particle spectra and effective hamiltonian}
\label{sec:formalism}

Spin-orbit coupled three-component Bose gas can be generalized
by counter-propagating Raman lasers along $\hat{x}$ to couple the
three hyperfine states with momentum transfer of the Raman process
$2k_{r}$. The non-interacting Hamiltonian can be written in the matrix
form as~\cite{lan-14}
\begin{equation}
\label{eqn:H0}
\mathcal{H}=\left(\begin{array}{ccc}
\frac{\left(k_{x}+2k_{r}\right)^{2}}{2}-\delta & \Omega_{R}/2 & 0\\
\\
\Omega_{R}/2 & \frac{ k_{x}^{2}}{2}-\epsilon & \Omega_{R}/2\\
\\
0 & \Omega_{R}/2 & \frac{\left(k_{x}-2k_{r}\right)^{2}}{2}+\delta
\end{array}\right)+\frac{ k_{\perp}^{2}}{2},
\end{equation}
where $k_{\perp}=\sqrt{k_{y}^{2}+k_{z}^{2}}$ is the transverse momentum,
$\delta$ is the two-photon detuning from the Raman resonance, $\epsilon$
is the quadratic Zeeman shift induced by the magnetic field along
$\hat{y}$ , and $\Omega_{R}$ represents the Rabi frequency of the
Raman transition. Notice that throughout the manuscript, we use the natural units
of $\hslash=m=1$, and define $k_{r}$ and the recoil energy $E_{r}=k_{r}^{2}/2$
as the units of momentum and energy, respectively. 

\begin{figure}
\begin{centering}
\includegraphics[width=0.48\columnwidth,height=0.34\columnwidth]{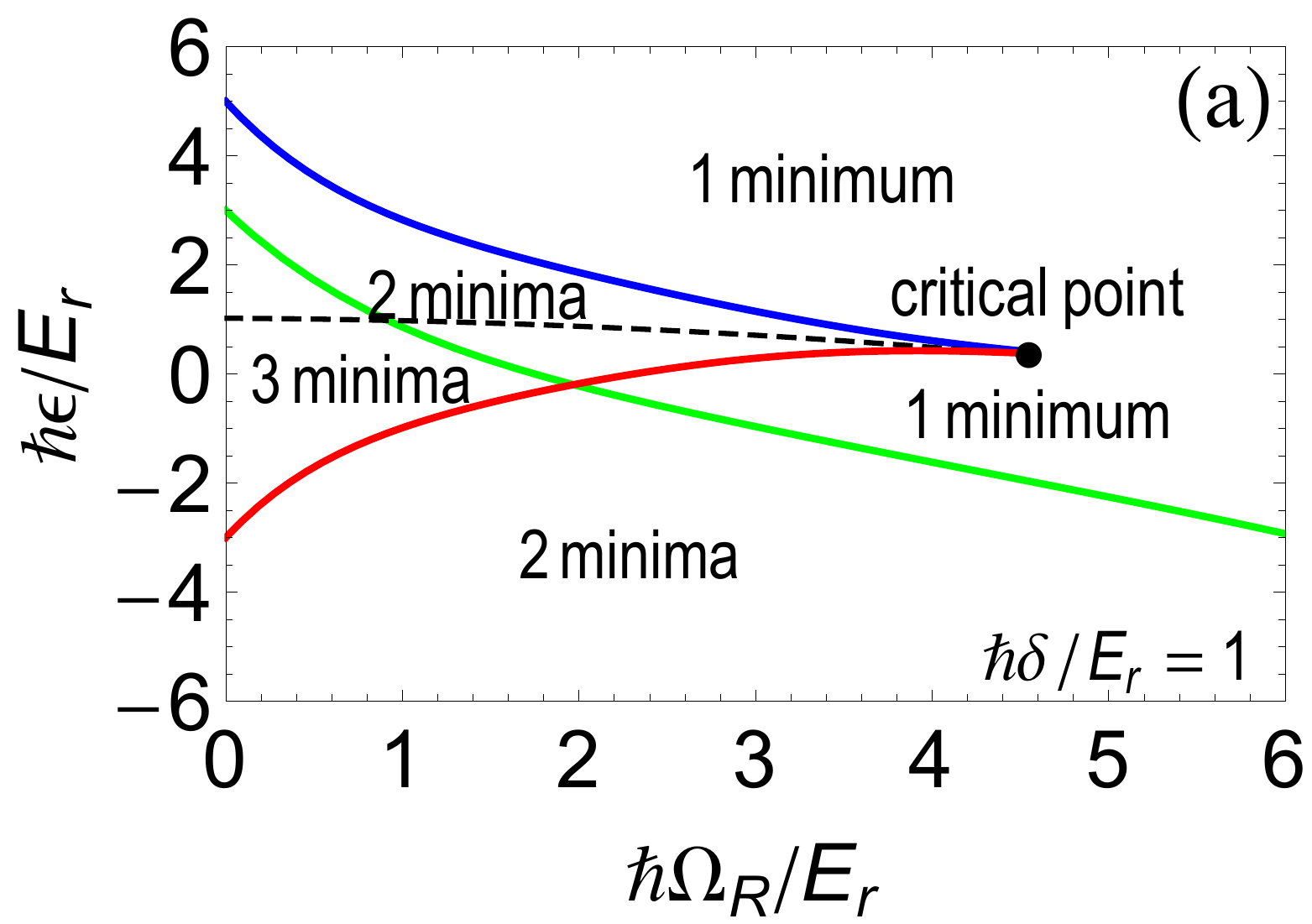}
\includegraphics[width=0.48\columnwidth,height=0.34\columnwidth]{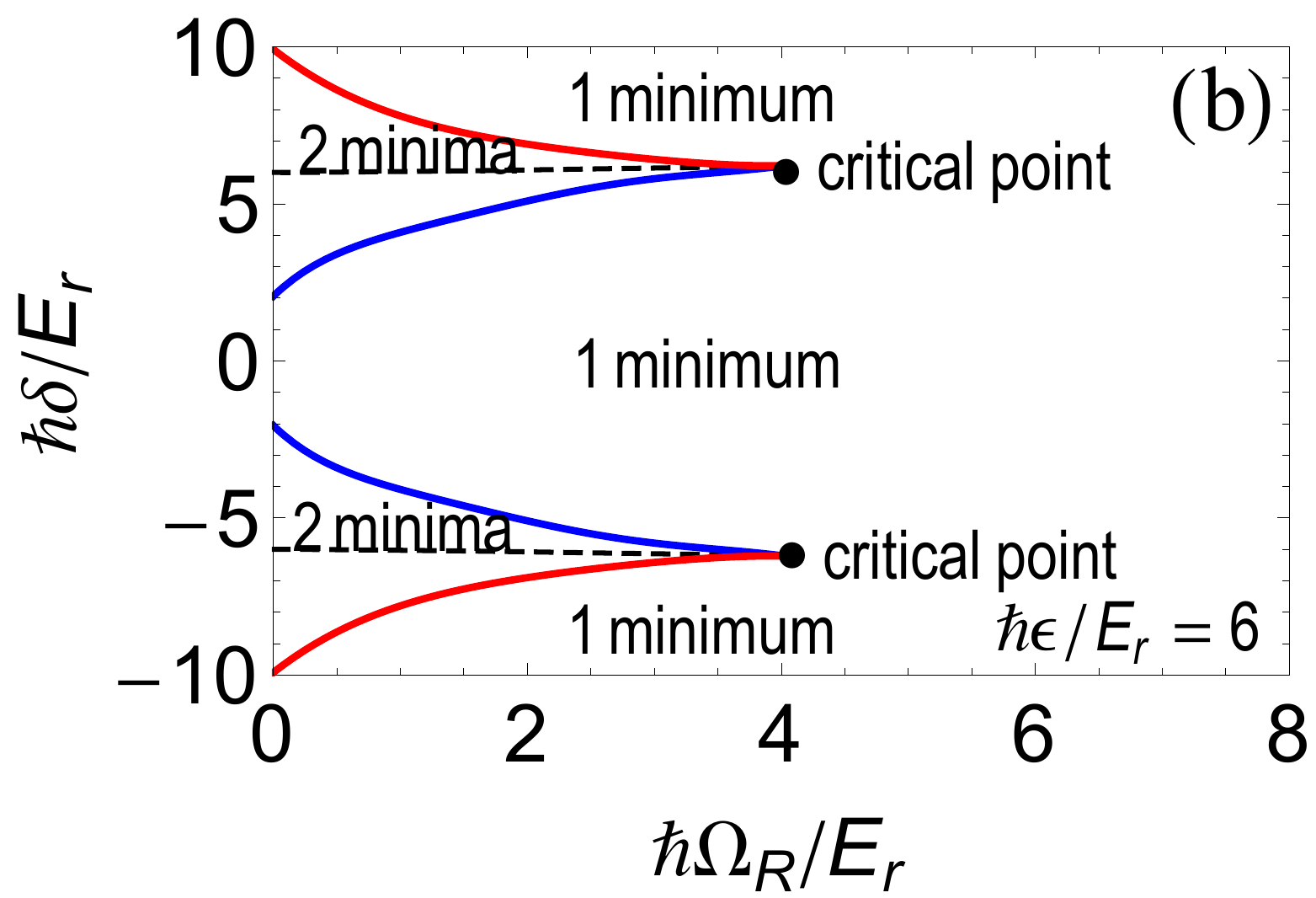}
\par\end{centering}
\centering{}
\includegraphics[width=0.48\columnwidth]{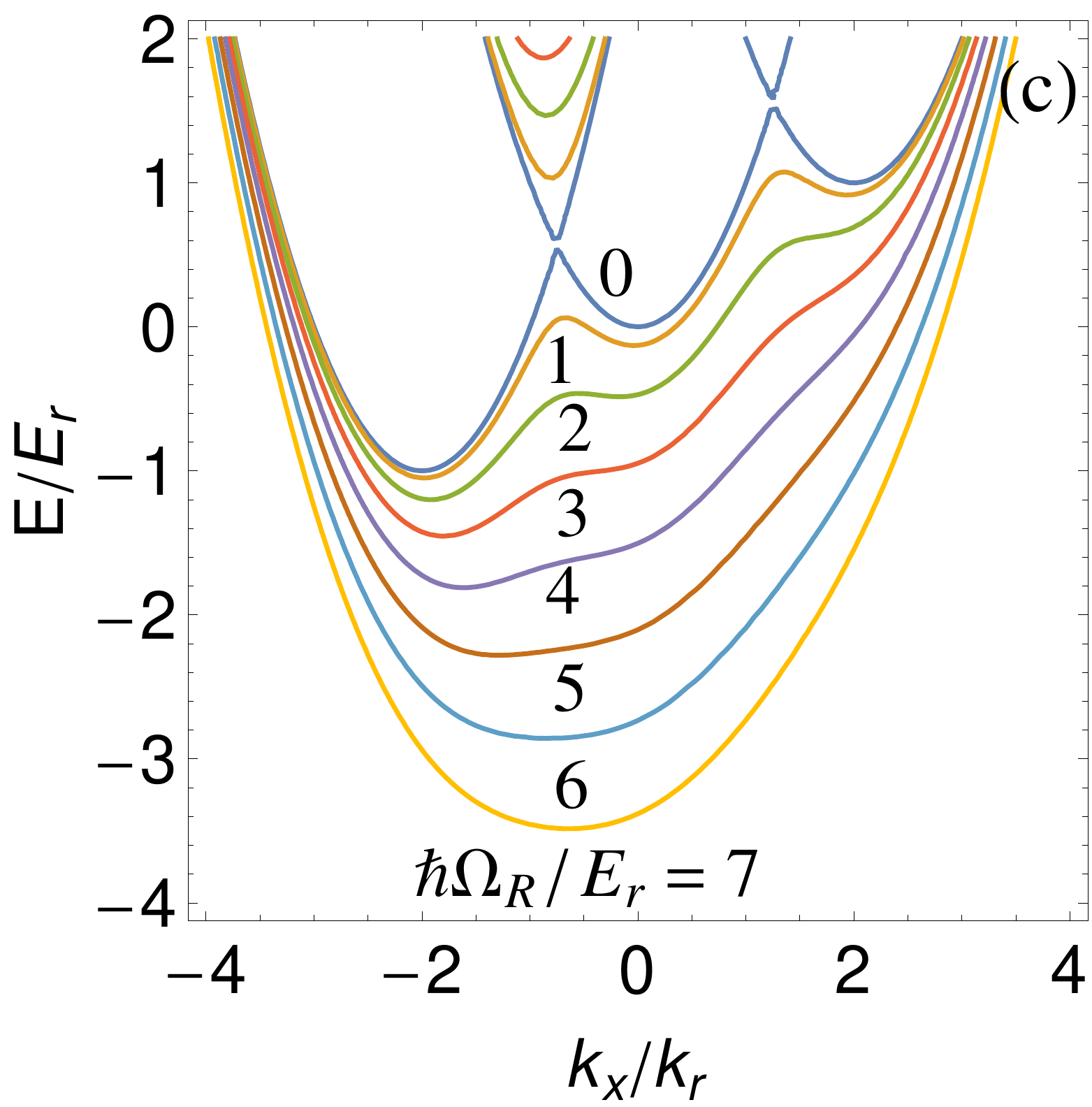} 
\includegraphics[width=0.48\columnwidth]{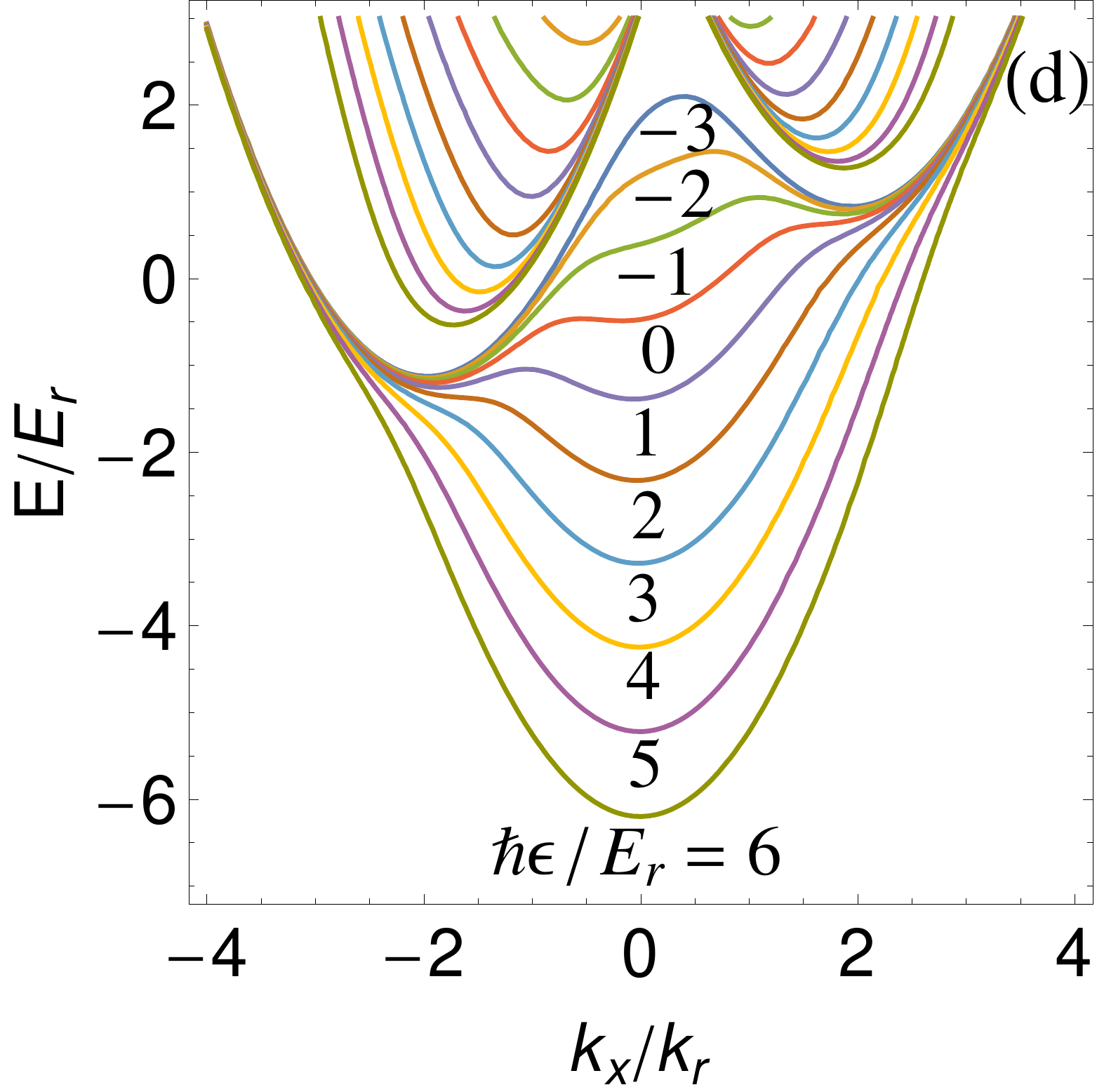}
\caption{(Color online) (a-b) Single-particle phase diagrams of a three-component
Bose gas with one-dimensional SOC in the (a) $\Omega_{R}$--$\epsilon$ 
plane with $\delta=1$ and (b) $\Omega_{R}$--$\delta$ plane with $\epsilon=6$.
The lowest branch of the single-particle dispersion spectrum acquires either 
one, two, or three local minima in different parameter regimes separated by solid lines. 
On the dashed lines within regions of multiple minima, two of the local minima 
are degenerate. Typical examples for the lowest branch of dispersion curves 
by changing (c) Rabi frequency $\Omega_{R}$ with $\delta=1$ and $\epsilon=0$ 
and (d) quadratic Zeeman energy $\epsilon$ with $\delta=1$ and $\Omega_{R}=2$.}
\label{fig:dispersion}
\end{figure}

The single-particle dispersion can be obtained by diagonalizing th non-interacting 
Hamiltonian of Eq. (\ref{eqn:H0}). The resulting spectra has three branches, among 
which the lowest one can have three minima, two minima, 
or one minimum depending on the combination of parameters. 
In Figs.~\ref{fig:dispersion}(a) and \ref{fig:dispersion}(b), we show the parameter 
regions exhibiting different structures for the case of $\hbar\delta/E_r=1$ and $\hbar\epsilon/E_r=6$ respectively.
From Fig.~\ref{fig:dispersion}(a), we can identify various regions where the lowest branch of 
single-particle dispersion acquires 1, 2, or 3 minima. Specifically, for the case of a large positive 
quadratic Zeeman splitting $\epsilon$, the $\vert 0 \rangle$ state is far detuned from the 
other two high-lying hyperfine states, so that the spectrum has only one minimum. On the other hand, 
if $\epsilon$ is large negative, the $\vert 0 \rangle$ state becomes the high-lying state and the system
essentially turn into a spin-1/2 Bose gas where the two $\vert \pm 1 \rangle$ spin components 
are spin-orbit coupled via virtual processes involving the $\vert 0 \rangle$ state. 
As a result, the single-particle dispersion can have either two or one minima, depending on the SOC intensity $\Omega_R$ and two-photon detuning $\delta$. 
For the case of intermediate $|\epsilon|$, all three hyperfine states 
are spin-orbit coupled and the shape of spectrum is sensitively dependent on all parameters. 
Typical examples of dispersion spectra along the $k_{x}$ axis showing one minimum, two minima, 
and three minima, as well as the trends of evolution depending on $\Omega_R$ and $\epsilon$ 
are illustrated in Figs.~\ref{fig:dispersion}(c) and \ref{fig:dispersion}(d), respectively.

As the spin operators in spin-1/2 systems all belong to the SU(2) group, 
those in spin-1 systems discussed here are elements in the SU(3) group. 
The SU(3) group is locally isomorphic to the O(8) group, which has eight linearly independent 
observables as generators. These generators can be grouped into two types, including three spin
vectors (or angular momentum operators) and five nematic tensors.
The irreducible matrix representations of these observables are given by~\cite{{Yukawa-13}}

\noindent \begin{center}
$J_{x}=\frac{1}{\sqrt{2}}\left(\begin{array}{ccc}
0 & 1 & 0\\
1 & 0 & 1\\
0 & 1 & 0
\end{array}\right)$, $J_{y}=\frac{i}{\sqrt{2}}\left(\begin{array}{ccc}
0 & -1 & 0\\
1 & 0 & -1\\
0 & 1 & 0
\end{array}\right)$, 
\par\end{center}

$ $
\noindent \begin{center}
$J_{z}=\left(\begin{array}{ccc}
1 & 0 & 0\\
0 & 0 & 0\\
0 & 0 & -1
\end{array}\right)$, $Q_{xy}=i\left(\begin{array}{ccc}
0 & 0 & -1\\
0 & 0 & 0\\
1 & 0 & 0
\end{array}\right)$, 
\par\end{center}

$ $
\noindent \begin{center}
$Q_{yz}=\frac{i}{\sqrt{2}}\left(\begin{array}{ccc}
0 & -1 & 0\\
1 & 0 & 1\\
0 & -1 & 0
\end{array}\right)$, $Q_{zx}=\frac{1}{\sqrt{2}}\left(\begin{array}{ccc}
0 & 1 & 0\\
1 & 0 & -1\\
0 & -1 & 0
\end{array}\right)$,
\par\end{center}

$ $
\noindent \begin{center}
$D=\left(\begin{array}{ccc}
0 & 0 & 1\\
0 & 0 & 0\\
1 & 0 & 0
\end{array}\right)$, $Y=\frac{1}{\sqrt{3}}\left(\begin{array}{ccc}
1 & 0 & 0\\
0 & -2 & 0\\
0 & 0 & 1
\end{array}\right).$
\par\end{center}
The commutators between these spin operators can then be classified into three
categories: $[J_{y},J_{z}]=iJ_{x}$ as spin-spin group, $[Q_{xy},Q_{xz}]=iJ_{x}$,
$[Q_{yz},D]=iJ_{x}$, and $[Q_{yz},Y]=\sqrt{3}iJ_{x}$ as nematic-nematic
group, and $[J_{x},Q_{yz}]=i(\sqrt{3}Y+D)$, $[J_{y},Q_{zx}]=i(-\sqrt{3}Y+D)$ 
as spin-nematic group.

To study the effective spin-spin interaction induced by SOC, as well as the induced 
spin squeezing effect, next we derive an effective spin model. To facilitate the derivation, we impose a weak harmonic trap 
$V(x)= \omega_{x}^{2} x^2 /2+ \omega_{y}^{2} y^2 /2+ \omega_{z}^{2} z^2 /2$. 
We will find that the resulting form of the effective model 
does not depend on the absolute value of trapping frequency, hence 
incorporate solely the effect of SOC. In the presence of such an auxiliary trapping potential, 
we can quantize the motional degrees of freedom along the trapping direction to 
a discrete energy spectrum. In particular, by introducing the bosonic operators
$a\equiv \sqrt{\omega_{x}/2}(x+ik_x/\omega_{x})$, $b\equiv \sqrt{\omega_{y}/2}(y+ik_y/\omega_{y})$, 
$c\equiv \sqrt{\omega_{z}/2}(z+ik_z/\omega_{z})$ and the collective spin operators 
$F_{s=x,y,z}=\sum_{i=1}^{N} J_{i,s}$, $F_{Y}= \sum_{i=1}^{N} Y_i$, 
the Hamiltonian of Eq.~(\ref{eqn:H0}) for N-particle can be rewritten as
\begin{eqnarray}
\tilde{H} & = & \omega_{x} Na^{\dagger} a+ N\frac{4 k_r^2 - \epsilon}{3}
+ \frac{\Omega_{R}}{\sqrt{2}} F_{x} \nonumber \\ 
&& + i k_{r} \sqrt{2 \omega_{x}} (a^{\dagger}-a) F_{z}
- \delta F_{z}  + \frac{ 2 k_{r}^{2} + \epsilon}{\sqrt{3}} F_{Y}.
\end{eqnarray}
Here we ignore $\omega_{y} Nb^{\dagger} b+\omega_{z} Nc^{\dagger} c$ since the boson modes in $y$,$z$ direction do not interact with the ultracold atoms. Employing the unitary transformation $U = \exp[ i G( a^{\dagger} +a)F_{z} ]$
with $G=\sqrt{2/\omega_{x}} k_{r}/N$, the Hamiltonian thus can be transformed as
\begin{eqnarray}
\label{eqn:H2}
\tilde{H}' & = & \omega_{x} Na^{\dagger}a - q F_{z}^{2} -\delta F_{z} 
+ \frac{2k_{r}^{2} + \epsilon}{\sqrt{3}} F_Y
\nonumber \\ 
& + & \frac{\Omega_{R}}{\sqrt{2}}  \{F_{x} \cos [ G(a^{\dagger}+a)]-F_{y} \sin [ G(a^{\dagger}+a)]\},
\end{eqnarray}
where $q=4k_r^2/N=8E_r/N$. Notice that the term of $N(4k_r^2 - \epsilon)/3$ has been dropped out as the zero-point energy.

For a BEC, the expectation value of $\langle a^\dagger a \rangle$ is in the order of $N$ for 
the ground state, and about unity for excited states. Considering the prefactor of $1/N$ in the definition
of $G$, the leading order of the arguments in the cosine and sine functions in Eq. (\ref{eqn:H2}) are 
$1/{\sqrt{N}}$, which is negligible for systems of large particle number. As a result, we can approximate 
the cosine and sine functions to the zeroth order, and the Hamiltonian Eq. (\ref{eqn:H2}) becomes 
separable in spatial and spin degrees of freedom, leading to an effective spin Hamiltonian
\begin{eqnarray}
\label{eqn:Heff}
H_{{\rm eff}}=-qF_{z}^{2}+\frac{\Omega_{R}}{\sqrt{2}}F_{x}-\delta F_{z}+\frac{4E_r+\epsilon}{\sqrt{3}}F_{Y}.
\end{eqnarray}
One can see clearly that an effective spin-spin interaction emerges as a result of SOC, and the Rabi frequency $\Omega_R$, two-photon detuning $\delta$, and the quadratic Zeeman splitting $\epsilon$ act as effective Zeeman fields along different directions in 
the eight-dimensional spin hyperspace.

\section{spin-nematic squeezing}
\label{sec:results}

With the aid of the effective spin model of Eq. (\ref{eqn:Heff}), we can study the spin squeezing in 
the underlying system. As the commutators relation between spin and nematic operators are not present 
in the spin-1/2 case, next we focus on spin squeezing of this type. The method can be straightforwardly 
applied to the spin-spin and nematic-nematic commutators, and the results are qualitatively consistent
with the findings for the spin-spin case in spin-1/2 system with SOC~\cite{chen-13, huang-15}.

The spin model of Eq. (\ref{eqn:Heff}) can not be solved analytically due to the presence of nonlinear 
interaction. In the low excitation limit, however, we can introduce the Holstein-Primakoff transformation 
for spin-1 systems
\begin{eqnarray}
F_{x} & \equiv & 
\frac{1}{\sqrt{2} } \left(b_{1}^{\dagger} N_0' + N_0' b_{-1} + \rm{h.c.} \right), 
\nonumber \\
F_{y} & \equiv & 
\frac{1}{\sqrt{2} i } \left( b_{1}^{\dagger} N_0' + N_0' b_{-1} - \rm{h.c.}\right),
\end{eqnarray}
where $N_0' \equiv \sqrt{N-b_{1}^{\dagger}b_{1}-b_{-1}^{\dagger}b_{-1}}$, and the operators 
$b_{1}$ and $b_{-1}$ representing spin flipping processes between the internal levels $| \pm 1 \rangle$ and $| 0 \rangle$, represented by the bosonic modes $a_{\pm1}$ and $a_0$. For the case that most of the particles remain in the mode $a_0$, i.e., $\langle a_0^\dagger a_0\rangle \simeq N$ and $\langle b_{\pm1}^\dagger b_{\pm1}\rangle\ll N$, the operators $b_{\pm1}=a_{\pm1}a_0^\dagger/\sqrt{N}$ are effective bosonic modes satisfying 
the bosonic commutation relations $\left[b_{\alpha},b_{\beta}^{\dagger}\right]=\delta_{\alpha\beta}$ 
with $\alpha,\beta=\pm1$. Within the assumption that the majority of the particles are residing in the 
$| 0 \rangle$ state, or equivalently the excitations to the $| \pm 1 \rangle$ states are rare, we can 
rewrite the bosonic operators as a mean-field value plus some fluctuations 
$b_{\pm1}=\sqrt{N}\beta_{\pm1}+\delta b_{\pm1}$.
The ground state energy can the be obtained by minimizing the energy functional $E(\beta_{1},\beta_{-1})$. As in this the low excitation limit, nearly all the spins are polarized in $F_{Y}$ direction, which means $\left| \left\langle \pm\sqrt{3}F_{Y}+F_{D} \right\rangle \right| \approx 2N$, 
the squeezing parameter is then given by~\cite{sun-11}
\begin{equation}
\label{eqn:xi}
\xi_{x}\equiv\frac{\min\left(\triangle^{2}J_{n_{\perp}}\right)}{J/2}\approx\triangle^{2}F_{x}/N,
\end{equation}
Here, $J$ is the expectation value of mean spin, $J_{n_\perp}$ is a spin component along the direction perpendicular to the mean spin direction. So in our case, it is clear that $\xi_{x}$ can be obtained by calculated the variance of $F_{x}$, one has spin squeezing in the spin-nematic channel as $\xi_x < 1$.

\begin{figure}
\centering{}\includegraphics[width=7.5cm]{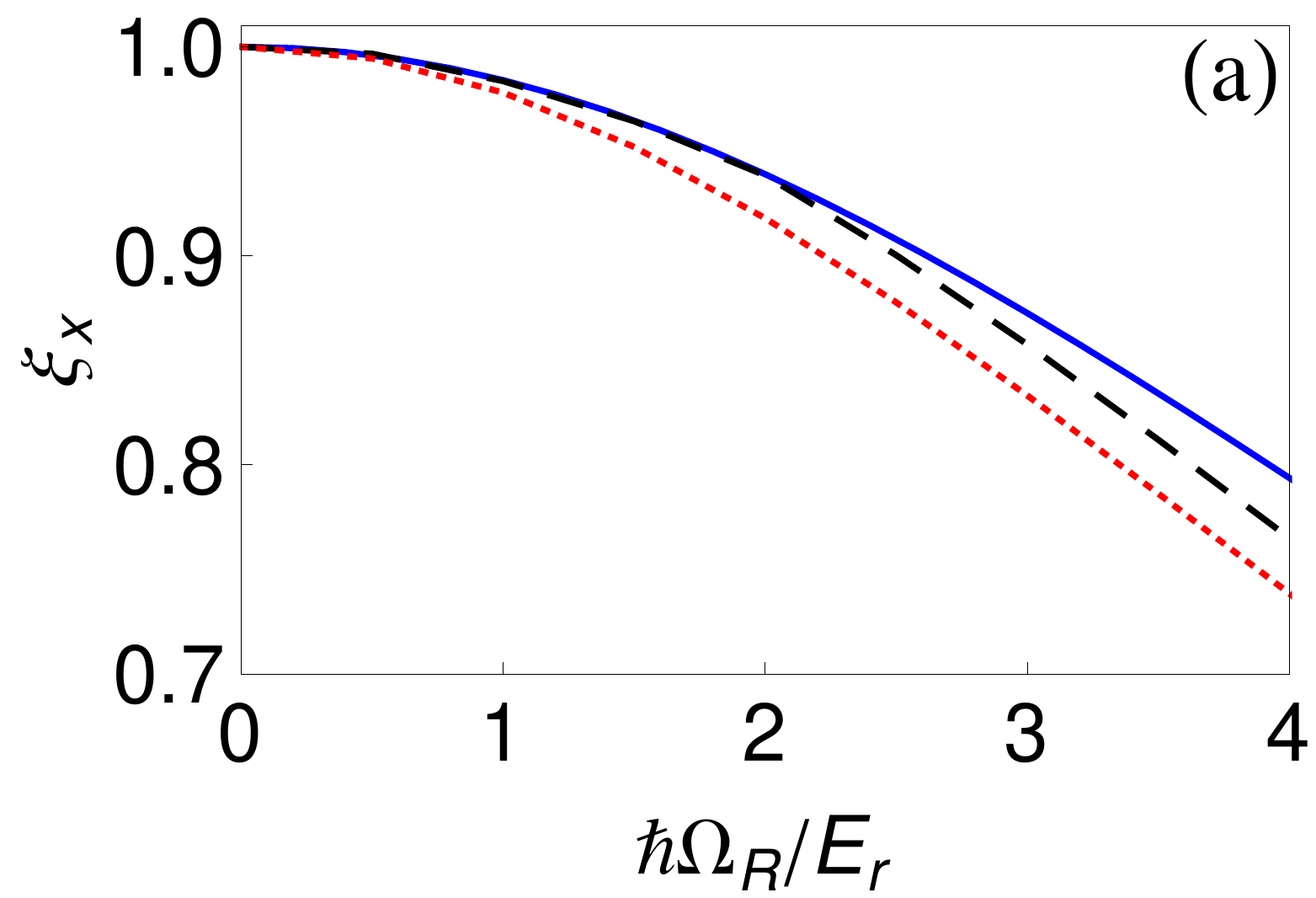}
\includegraphics[width=7.7cm]{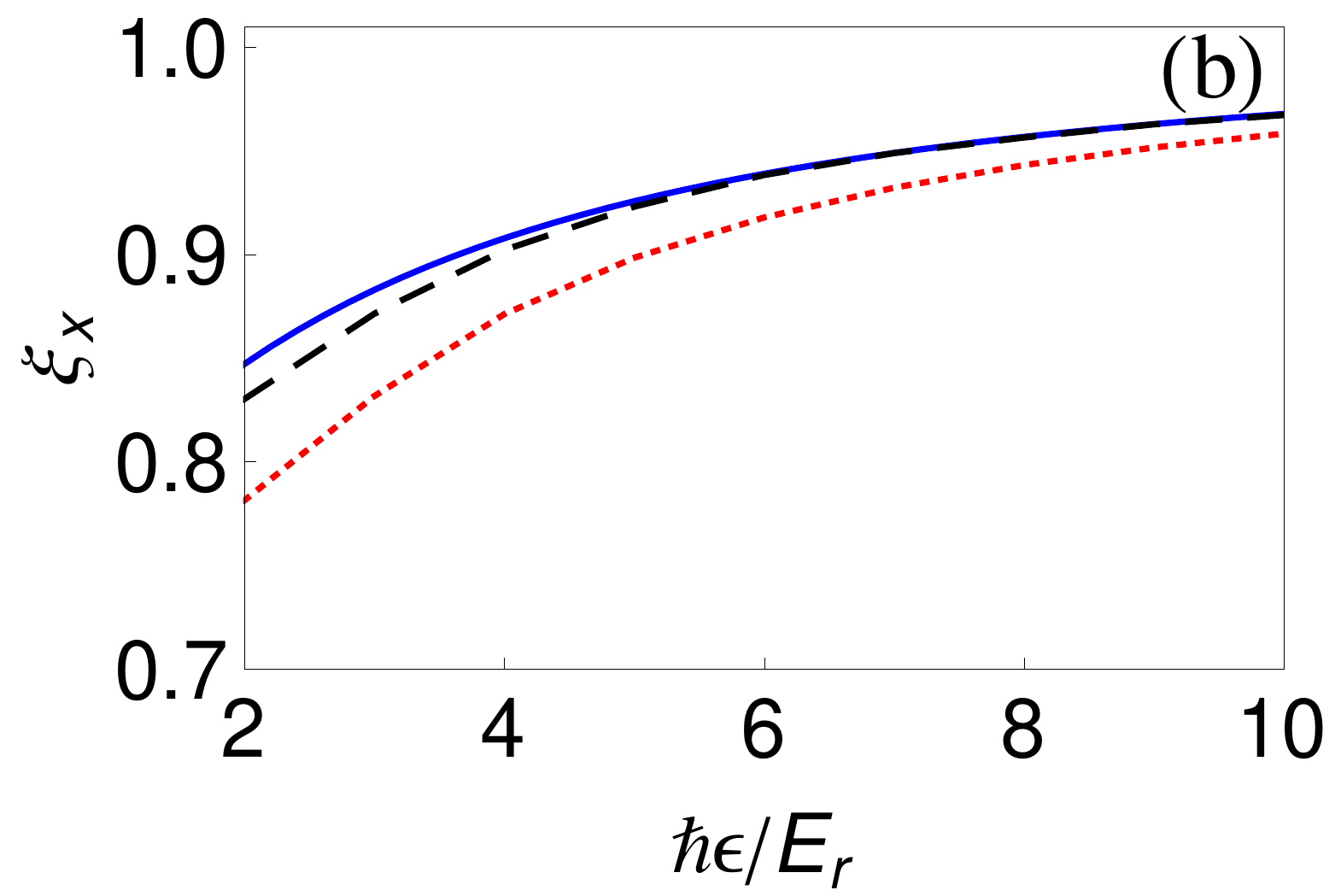}
\caption{(Color online) Spin-nematic squeezing parameter $\xi_{x}$ as a function
of (a) Rabi frequency $\Omega_{R}$ with $\delta=0$ and $\epsilon=6$
and (b) quadratic Zeeman splitting $\epsilon$ with $\delta=0$ and $\Omega_{R}=2$.
In both figures, results obtained from the effective spin model Eq.(\ref{eqn:Heff}) 
are illustrated by blue solid lines, in comparison to the numerical solutions of the GP 
equation for a pancake-shaped trap with $\omega_{x}=\omega_{y}=50$ Hz, 
$\omega_{z}=1500$ Hz (black dashed), and for a cigar-shaped trap with 
$\omega_{x}=\omega_{y}=5000$ Hz, $\omega_{z}=1500$ Hz (red dotted). 
Here, we consider a gas of $^{87}$Rb atoms in the $F=1$ manifold with background 
interaction and total particle number $N=10^5$.}
\label{fig:squeezing1}
\end{figure}

We first discuss the case of zero two-photon detuning $\delta = 0$, and show 
in Fig.~\ref{fig:squeezing1} the spin-nematic squeezing parameter as functions of 
Rabi frequency $\Omega_{R}$ and quadratic Zeeman splitting $\epsilon$.
One can see clearly that the ground state is a spin squeezed state
under the effect of SOC. Importantly, as shown in Fig.~\ref{fig:squeezing1}(a),
spin-nematic squeezing can be enhanced with increasing $\Omega_{R}$.
This behavior is in stark contrast to the case of spin-1/2 systems, where the spin-spin 
squeezing is favored by decreasing $\Omega_R$~\cite{chen-13, huang-15}.

\begin{figure}[t]
\includegraphics[width=7.5cm]{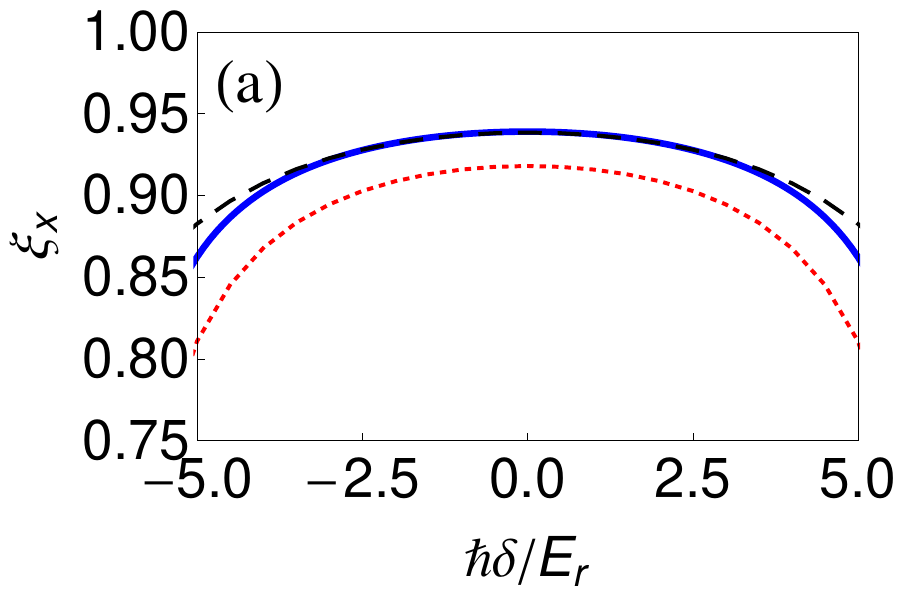}
\includegraphics[width=7.5cm]{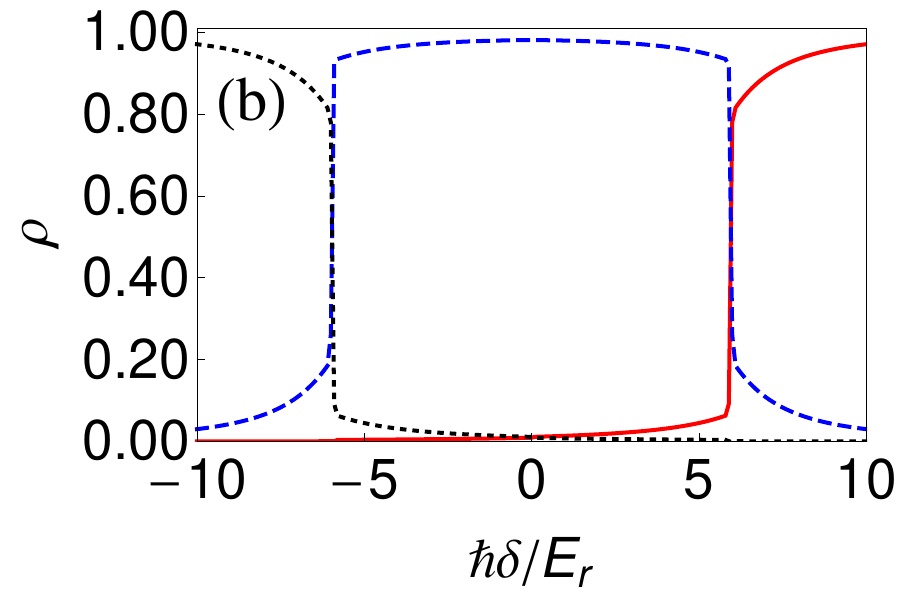}
\caption{(Color online)(a) Variations of spin-nematic squeezing parameter $\xi$ 
as a function of two-photon detuning $\delta$ with $\Omega_{R}=2$ and $\epsilon=6$. 
Analytic result obtained from the effective spin model Eq. (\ref{eqn:Heff}) within low-density
excitation approximation (blue solid) is compared with numerical solutions of the GP 
equation for a pancake-shaped trap with $\omega_{x}=\omega_{y}=50$ Hz, 
$\omega_{z}=1500$ Hz (black dashed), and for a cigar-shaped trap with 
$\omega_{x}=\omega_{y}=5000$ Hz, $\omega_{z}=1500$ Hz (red dotted). 
(b) Atom number fractions of the $| -1\rangle$ (black dotted), $| 0\rangle$ (blue dashed), 
and $| +1 \rangle$ (red solid) states.}
\label{fig:squeezing2}
\end{figure}

\vspace{-1em}
We then extend the discussion to the more general case of a nonzero
two-photon detuning $\delta\neq0$. This scenario is experimentally
relevant because a severe heating effect is usually present as the
Raman transition is on-resonance. As shown in Fig.~\ref{fig:squeezing2}(a), 
a finite $\delta$ favors spin-nematic squeezing within a fairly large region
of $\left|\hbar\delta/E_r\right|<5$. This result can be understood by analyzing the 
single-particle Hamiltonian of Eq.~\ref{eqn:H0}, where $\delta$ and $\epsilon$ 
are energy offsets of the diagonal elements. As Raman transitions will be 
enhanced when difference states are near resonance, spin squeezing will be 
favored when the absolute value of $\delta$ is close to $\epsilon$. 
To further clarify this argument, we analyze the atom populations of different 
ground states with changing $\delta$. As shown in Fig.~\ref{fig:squeezing2}(b),
the presence of a finite $\delta$ will enhance the transition between 
the $| 0 \rangle$ state and one of the $| \pm 1 \rangle$ states, while the transition 
to the other $| \pm 1 \rangle$ state is reduced. Notice that this behavior is 
very different from the spin-1/2 case, where the two spin components are moved 
away from each other with increasing $\delta$, leading to an effectively weaker SOC. 

The dependences of spin-nematic squeezing on the various parameters of $\Omega_R$, 
$\epsilon$ and $\delta$ can also be interpreted from the effective spin model of Eq. (\ref{eqn:Heff}), 
within which the three parameters correspond to effective Zeeman fields along the $F_x$, $F_Y$, 
and $F_y$ directions, respectively. Considering that in the low excitation limit nearly all spins 
are polarized along the $F_Y$ direction, a stronger Zeeman field along the same direction, i.e., 
a larger value of $\epsilon$, will further intensify the polarization so that the effective spin-spin interaction 
becomes relatively weak, leading to a less spin-nematic squeezing effect. On the other hand, 
effective Zeeman fields along the perpendicular directions, either $F_x$ or $F_z$, will tilt the spin
polarization from the $F_Y$ axis slightly but the effect spin-spin interaction is enhanced obviously, resulting 
an increased squeezing parameter as in Eq. (\ref{eqn:xi}).
\begin{figure}[t]
\centering{}
\includegraphics[width=7.5cm]{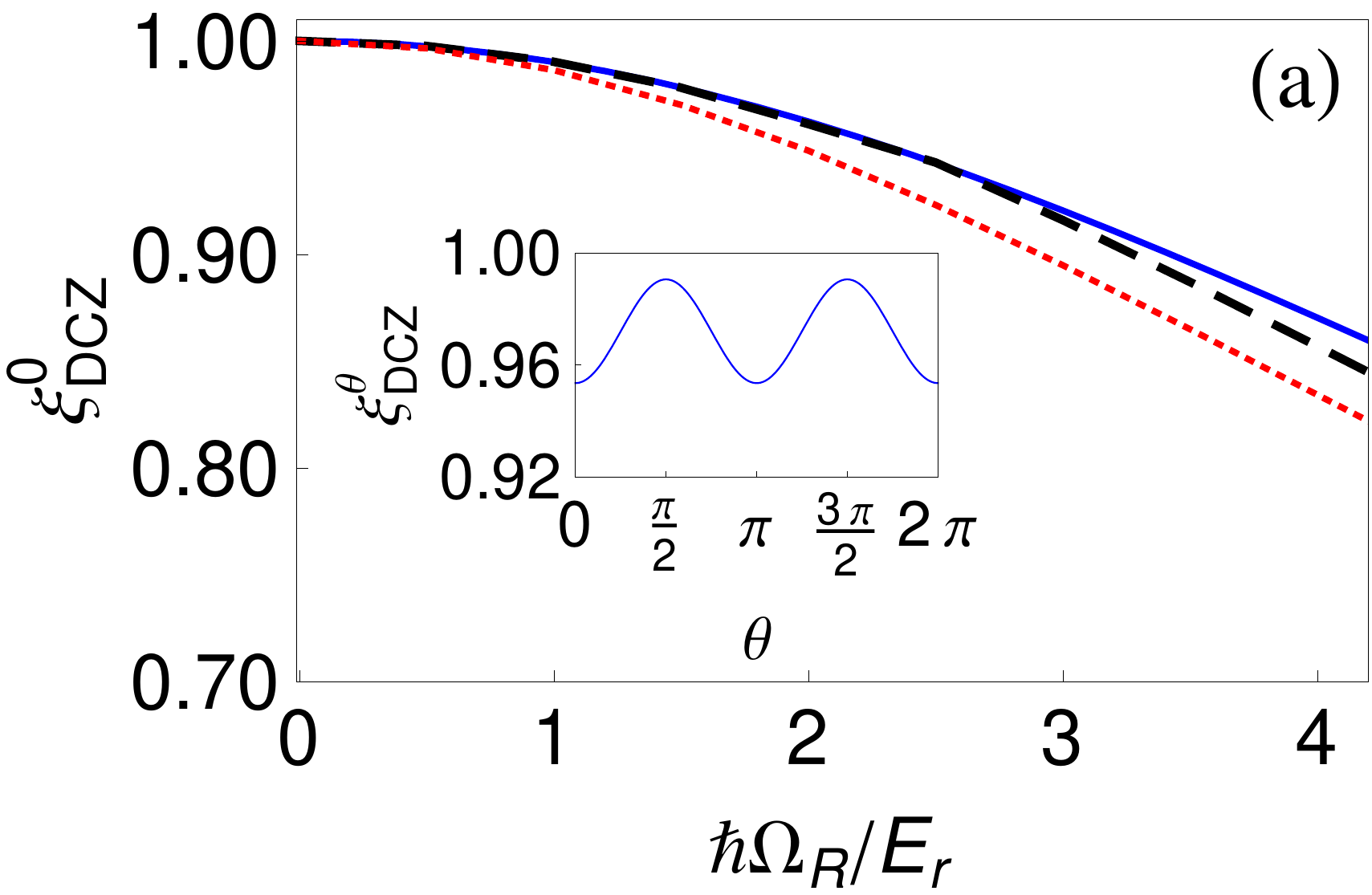}
\includegraphics[width=7.5cm]{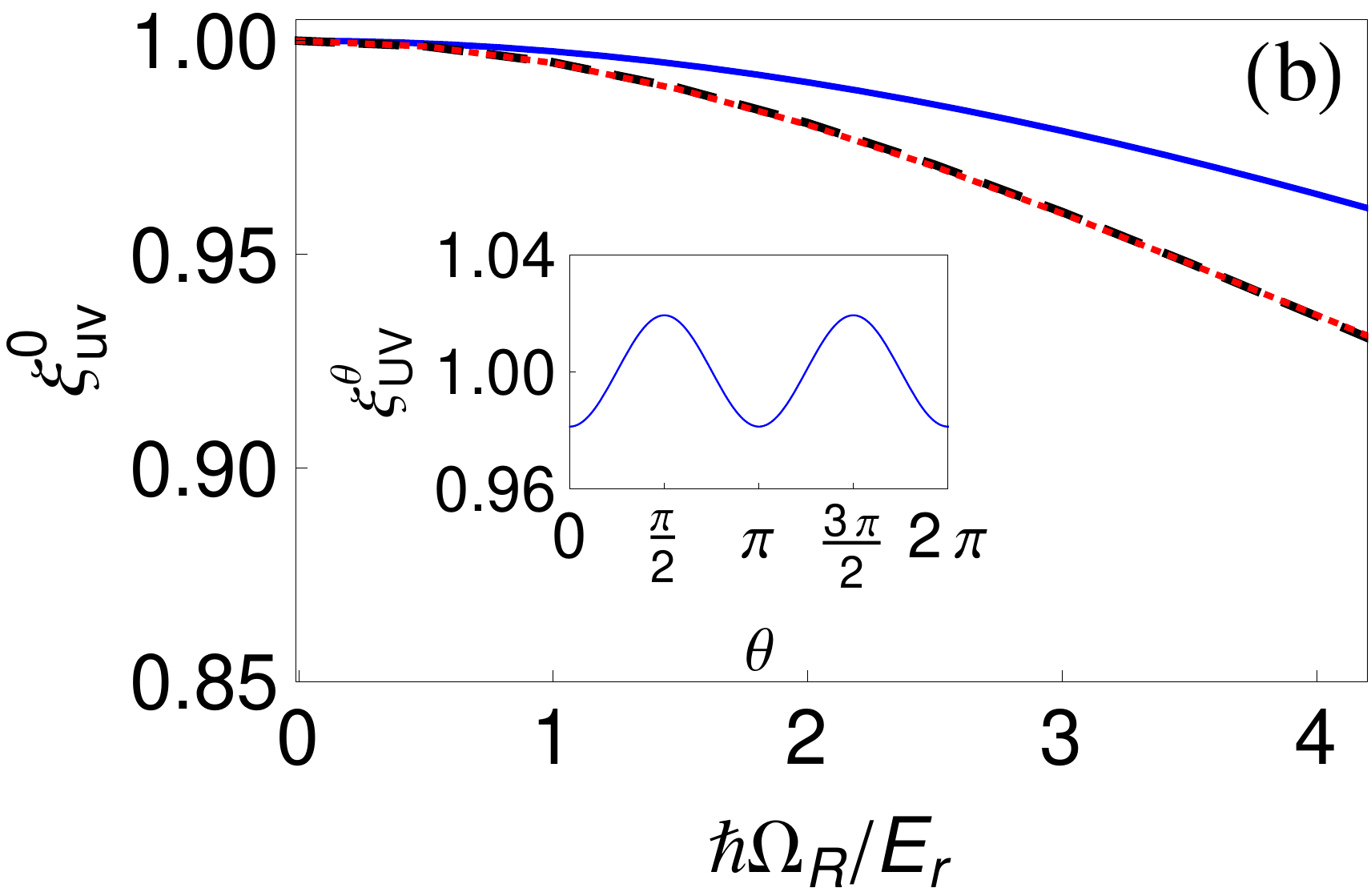}
\caption{(Color online) (a) Two-mode entanglement parameter $\xi_{DCZ}^{0}$ and (b) two-spin
squeezing parameter $\xi_{UV}^{0}$ versus Rabi frequency $\Omega_{R}$ with other 
parameters being $\delta=0$ and $\epsilon=6$. 
Analytic result obtained from the effective spin model Eq. (\ref{eqn:Heff}) within low-density
excitation approximation (blue solid) is compared with numerical solutions of the GP 
equation for a pancake-shaped trap with $\omega_{x}=\omega_{y}=50$ Hz, 
$\omega_{z}=1500$ Hz (black dashed), and for a cigar-shaped trap with 
$\omega_{x}=\omega_{y}=5000$ Hz, $\omega_{z}=1500$ Hz (red dotted).
The insets show the squeezing parameters as functions of $\theta$. 
Notice that the optimal squeezing in both criteria are obtained when $\theta=n\pi$ with $n$ an integer. 
}
\label{fig:entanglement}
\end{figure}

In addition to the spin-nematic squeezing, we notice that in the low excitation limit 
with the majority of particles residing in the $| 0 \rangle$ state, the two effective 
bosonic modes $b_{1}$ and $b_{-1}$ can be entangled, which is referred as 
two-mode entanglement. A sufficient criterion for entanglement between the modes 
$b_{1}$ and $b_{-1}$ from the spin squeezing parameters is then given by~\cite{duan-02}
\begin{eqnarray}
\label{eqn:dcz}
\xi_{{\rm DCZ}}^{\theta}=(\xi_{+}^{\theta}+\xi_{-}^{\theta+\pi/2})/2<1,
\end{eqnarray}
where $\xi_{\pm}^{\theta}\approx\langle\Delta^{2}F_{\pm}^{\theta}\rangle/N$ represents the variance 
of quadrature phase amplitudes which depends on the parameter $\theta$, and we have use the definitions $F_{+}^{\theta}=\cos\theta F_{x}+\sin\theta F_{yz}$ and $F_{-}^{\theta}=\cos\theta F_{zx}+\sin\theta F_{y}$ in this system. Here, the collective nematic operators are $F_{yz} =\sum_{i=1}^N Q_{yz}$ and $F_{zx} =\sum_{i=1}^N Q_{zx}$.
%
Figure~\ref{fig:entanglement}(a) shows that $\xi_{{\rm DCZ}}^{\theta}$ reaches
its minimum for $\theta=n\pi$ with $n$ an integer, and the entanglement is enhanced by Raman transition. 

Another representation of two-mode entanglement is called two-spin squeezing,
which is defined by dividing the spin-1 space into three subspaces pseudospins (each of spin-1/2)  $U$, $V$ and $T$ associated the three relative number differences of particles $N_{+1}-N_{0}$, $N_{-1}-N_{0}$, and $N_{+1}-N_{-1}$ in three-component labeled by $\{+1,-1,0\}$ ~\cite{lyou-02}. Two-spin squeezing parameter is given to describe the correlation between the spin subspace $U$ (the spin flipping process between internal levels $|+1\rangle$ and $|0 \rangle$) and $V$ (the spin flipping process between internal levels $|-1\rangle$ and $|0 \rangle$)~\cite{lyou-02}
\begin{eqnarray}
\label{eqn:uv}
\xi_{{\rm UV}}^{\theta}=\frac{\Delta^{2}F_{+}^{\theta}+\Delta^{2}F_{-}^{\theta+\pi/2}}{\sqrt{3}|\langle F_Y\rangle|}<1,
\end{eqnarray}
%
In Fig.~\ref{fig:entanglement}(b), one can see clearly that the optimal correlation 
is obtained when $\theta=n\pi$ with $n$ an integer, and increases with the the Raman transition. 
When comparing spin-nematic squeezing parameter (Fig.~\ref{fig:squeezing1}(a)) with these two criterions (Fig.~\ref{fig:entanglement}), we find that the effect of squeezing in spin-nematic channel is another representation of the correlation between two spin subspaces and entanglement between two effective modes in the low excitation limit.

Finally, we notice that in realistic experiments, one also needs to take the effects of inter-atomic
interaction and a global harmonic trap into consideration. Taking
$^{87}$Rb as a particular example, the interaction among the three
hyperfine states of the ground state manifold can be categorized into
two groups, depending on the total angular momentum of the two colliding
atoms. The background scattering lengths are taken as $a_{s0}=101.8a_{0}$
for $F=0$, and $a_{s2}=100.4a_{0}$ for$F=2$, where $a_{0}$ denotes
the Bohr radius~\cite{Zhang-05}. For the effects of trapping potentials, we consider 
two types of global harmonic traps including a pancake-shaped
quasi-two-dimensional trap with $\omega_{x}=\omega_{y}=50$ Hz and
$\omega_{z}=1500$ Hz, and a cigar-shaped three-dimensional trap with
$\omega_{x}=\omega_{y}=5000$ Hz and $\omega_{z}=1500$ Hz. 

By numerically solving the Gross-Pitaevski (GP) equation for a total number of 
$N=10^5$ atoms, we obtain the ground state of the system, and calculate the 
spin-nematic squeezing parameter $\xi_x$, the two-mode entanglement parameter 
$\xi_{\rm DCA}^\theta$, and the two-spin squeezing parameter $\xi_{\rm UV}^\theta$.
The corresponding results are shown in Figs.~\ref{fig:squeezing1}, \ref{fig:squeezing2} 
and \ref{fig:entanglement}. By comparing the numerical results with the outcome
from the effective spin model, we conclude that the effective model Eq. (\ref{eqn:Heff}) 
is qualitatively valid in the low excitation limit. On the other hand, a strong harmonic trap 
can cause sizable increment on spin-nematic squeezing and two-mode entanglement. 
This observation can be understood by noticing that in the presence of a strong harmonic trap, 
the particles will be more condensed with a higher number density at the trap center. 
As a result, the inter-particle interaction has stronger effect and causes better 
spin-nematic squeezing and two-mode entanglement.

\section{experimental detection and conclusion}
\label{sec:conclusion}

We have shown that an effective spin-spin interaction can be induced
in spin-orbit coupled spin-1 BEC, which can produce a special kind
of squeezing called spin-nematic squeezing. This type of spin squeezing 
can be enhanced by increasing Raman transition intensity and decreasing 
quadratic Zeeman splitting. More importantly, the squeezing is favored by 
a finite two-photon detuning in a fairly large parameter regime, which could be 
beneficial for experiments to reduce heating effect. These behaviors 
are in clear contrast to the spin squeezing within spin-orbit coupled spin-1/2 systems,
where the trends of dependence on Raman transition intensity and two-photon 
detuning are opposite. We also observe SOC induced two-mode entanglement and 
two-spin squeezing in such a system, and investigate their dependence on Raman 
transition intensity. We further analyze the effects of inter-particle interaction and 
external harmonic trap by numerically solving the GP equation, and find good 
agreement with approximate solutions of the effective spin model.

In order to detect such an exotic type of spin squeezing in this system, one may need 
to rotate $J_{x}$ into the easily measured $J_{z}$ direction by applying a $\pi/2$ 
radio-frequency (RF) rotation about the $J_{y}$ axis. This operation can be accomplished 
with a two-turn coil on the experimental $y$-axis driven at the frequency 
splitting of the $m_{F}$ states. Then, we can measure the variance of spin 
via a spin-selective imaging technique.

\acknowledgments
This work is supported by NSFC (11274009, 11622428, 11274025, 11434011, 11522436, 61475006, and 61675007), NKBRP (2013CB922000) and the Research Funds of Renmin University of China (10XNL016, 16XNLQ03).


\end{document}